\documentclass[prl,twocolumn,superscriptaddress,aps]{revtex4-1}
\usepackage{empheq}
\usepackage{tabularx}
\usepackage{dsfont}
\usepackage{graphicx}
\usepackage[hidelinks]{hyperref}
\usepackage{stackrel}
\usepackage{stix} 
\usepackage{amsmath,amssymb,latexsym,mathrsfs}
\usepackage{natbib}
\usepackage{amsfonts}
\usepackage{srcltx}
\usepackage{color}
\usepackage{cancel}
\usepackage{xr}
\usepackage{placeins}

\usepackage{pifont}
\newcommand{\cmark}{\ding{51}}%
\newcommand{\xmark}{\ding{55}}%

\newcommand{\vdplus}{VD$^{+}$}

\allowdisplaybreaks

\newcolumntype{Y}{>{\centering\arraybackslash}X}
\newcolumntype{L}[1]{>{\raggedright\let\newline\\\arraybackslash\hspace{0pt}}m{#1}}
\newcolumntype{C}[1]{>{\centering\let\newline\\\arraybackslash\hspace{0pt}}m{#1}}
\newcolumntype{R}[1]{>{\raggedleft\let\newline\\\arraybackslash\hspace{0pt}}m{#1}}

\def\0{\emptyset}

\begin{document}

\title{A violation of the Harris-Barghathi-Vojta criterion}
\author{Manuel Schrauth}
\author{Jefferson S. E. Portela}
\author{Florian Goth}
\affiliation{Institute of Theoretical Physics and Astrophysics,	University of W\"urzburg, 97074 W\"urzburg, Germany}

\begin{abstract}
In 1974, Harris proposed his celebrated criterion: Continuous phase transitions in $d$-dimensional systems are stable against quenched spatial randomness whenever $d\nu>2$, where $\nu$ is the clean critical exponent of the correlation length. Forty years later, motivated by violations of the Harris criterion for certain lattices such as Voronoi-Delaunay triangulations of random point clouds, Barghathi and Vojta put forth a modified criterion for topologically disordered systems: $a\nu >1$, where $a$ is the disorder decay exponent, which measures how fast coordination number fluctuations decay with increasing length scale. Here we present a topologically disordered lattice with coordination number fluctuations that decay as slowly as those of conventional uncorrelated randomness, but for which the universal behaviour is preserved, hence violating even the modified criterion. 
\end{abstract}

\maketitle	

Spatial disorder arising from impurities, defects, and other inhomogeneities is intrinsic to most experimental realizations. Moreover, disorder, in the form of random lattices, provides a useful theoretical tool for the discretization of non-trivial spaces in fields ranging from classical equilibrium statistical systems to quantum gravity~\cite[{and references therein}]{janke2002}. It is therefore important from both theoretical and experimental standpoints to understand the influence of quenched randomness. In particular, with respect to the critical behaviour of physical systems, the fundamental question is whether disorder affects the nature of phase transitions, and a number of arguments and criteria have been offered.

The original relevance criterion, proposed by Harris~\cite{harris1974,harris2016}, states that a second-order transition in a $d$-dimensional system fulfilling the equation $d\nu>2$, where $\nu$ is the clean critical exponent of the correlation length, is stable against quenched spatial disorder. A relevant example is the 2D contact process (CP), for which that inequality is violated. When placed on a regular lattice with random dilutions, the CP phase transition is controlled by an infinite-randomness critical point, where the usual algebraic scaling relations are replaced by a logarithmic time evolution~\cite{vojta2005,vojta2009}. However, a different behaviour emerges when the CP is placed on another kind of spatially disordered lattice, namely a Voronoi-Delaunay (VD) construction, where any two of the (randomly distributed) sites are connected whenever their Voronoi (or proximity) cells share a common edge. Here, clean universal properties have been verified~\cite{oliveira2008a}, in clear contradiction with Harris' criterion. In order to address these puzzling results, Barghathi and Vojta~\cite{barghathi2014} refined the original Harris criterion in terms of spatial anti-correlations in the coordination number fluctuations, offering an explanation for the preserved universality of VD lattices. This modified condition, which we call Harris-Barghathi-Vojta (HBV) criterion, posits that quenched topological disorder is irrelevant in systems that satisfy
\begin{align}
a\nu >1,
\label{eq:HBV_criterion}
\end{align}
where $ a $ is the dimension dependent \emph{disorder decay exponent}, defined by the relation
\begin{align}
\sigma_Q \sim L_b^{-a}.
\label{eq:definition_decay_exponent}
\end{align}
In Eq.~\eqref{eq:definition_decay_exponent}, the lattice is subdivided into spatial blocks of length $L_b$ and $\sigma_Q$ is the standard deviation of the block-averaged coordination numbers from the asymptotic mean value. 
In Ref.~\cite{barghathi2014} it is also shown that in 2D, $a\!=\!1$ for uncorrelated disorder (e.g., in regular lattices with random dilutions), whereas $a=3/2$ for VD lattices. Hence, the modified criterion Eq.~\eqref{eq:HBV_criterion} successfully explains earlier results of the Ising model and the contact process on those geometries (see Table~\ref{tab:HBV_predictions} for a compilation of predictions and observations in 2D).

\begin{figure}[t]
	\centering
	\includegraphics[width=0.49\columnwidth]{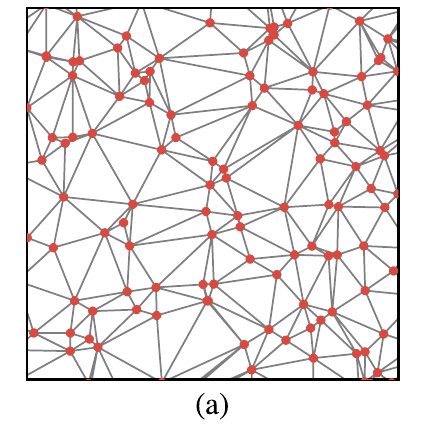}
	\includegraphics[width=0.49\columnwidth]{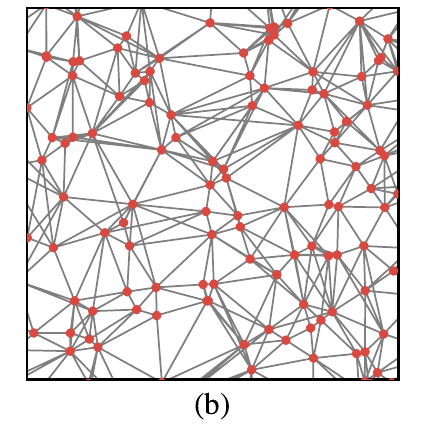}
	\caption{(a) Sample of a usual VD triangulation of a Poissonian point cloud. (b) The lattice we are introducing in this letter, \vdplus{}, constructed from (a) by adding random local bonds.}
	\label{fig:lattices}		
\end{figure}

\begin{figure}[t]
	\centering
	\includegraphics[width=0.9\columnwidth]{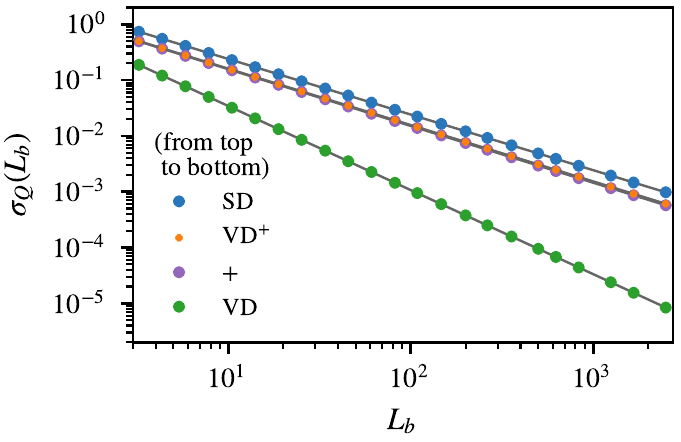}
	\caption{Coordination number fluctuations on different length scales for \vdplus{}, ordinary VD, the set of additional bonds that distinguishes both ($+$), and a site-diluted square lattice (SD). The curves are found to decay as $\sigma_Q \sim L_b^{-3/2}$ for the Voronoi-Delaunay triangulation (VD), and as $\sigma_Q \sim L_b^{-1}$ for the other lattices. For the analysis, 100 independent realizations of size $L=5000$ have been used each. Note that the data points for \vdplus{} and ($+$) almost coincide.}
	\label{fig:blocking}		
\end{figure}

\begin{table}[b]
	\small
	\centering
	\caption{Predictions of the HBV criterion for the 2D Ising and DP univerality classes as well as observations from numerical studies.}
	\begin{tabularx}{\columnwidth}{C{5mm}C{12mm}C{12mm}C{13mm}C{15mm}L{22mm}} 
		\hline
		\hline
		$ d $ & Class & Lattice & $a\nu $ & Prediction& Observation \\ 
		\hline		
		2 & Ising & VD & 3/2           & \cmark     & \cmark  $\,$ \cite{janke1993, janke1994,lima2000} \\ 
		2 & Ising & Diluted & 1             & -           &  marginal \cite{dotsenko1981,jug1983,shankar1987,shalaev1994,ludwig1987,kim1994a,kim1994b,selke1994b,ziegler1994,kuehn1994,kim2000,gordillo2009,martins2007,fytas2010,fytas2013,zhu2015}\\
		2 & DP    & VD & $ 1.100 $ & \cmark     & \cmark  $\,$ \cite{oliveira2008a,oliveira2016}\\
		2 & DP    & Diluted & $ 0.733 $ & \xmark &  \xmark  $\,$ \cite{vojta2005,vojta2006,vojta2009}\\ 
		2 & DP    & \vdplus{} & $ 0.733 $ & \xmark &  \cmark $\,$ [this work]\\ 
		\hline			
		\hline
	\end{tabularx} 
	\label{tab:HBV_predictions}
\end{table}

In this letter we define the \vdplus{} lattice: a Voronoi-Delaunay triangulation of a Poissonian point cloud, furnished with additional local bonds, see Fig.~\ref{fig:lattices}. This lattice is constructed from a VD triangulation of $N$ sites, to which $kN$ bonds between next-nearest neighbours are randomly added (we select $k=1$), resulting in a lattice with a total coordination number of exactly $(6+2k)N$. This latter constraint is relevant for the applicability of the HBV criterion~\cite{barghathi2014}\footnote{Note that, from geometrical constraints, any triangular lattice with $N$ vertices on a torus has exactly $6N$ bonds~\cite{barghathi2014}}.

We find the disorder decay exponent for the \vdplus{} to be  $a=1$ (see Fig.~\ref{fig:blocking}). This follows from the additional bonds being a source of uncorrelated disorder, which decays more slowly than the coordination number fluctuations of the original VD lattice. This interpretation is corroborated by the measurement of $\sigma_Q(L_b)$ for the additional bonds alone: as can be seen in Fig.~\ref{fig:blocking}, both curves are almost identical, demonstrating the dominance of the additional bonds. The HBV criterion therefore predicts that our construction should display a non-universal behavior for any universality class with $\nu<1$, such as directed percolation. In order to verify this prediction, we perform extensive numerical simulations of the contact process on the 2D \vdplus{} lattice and find strong indication of clean universal behavior, contradicting the HBV prediction.

The contact process is a simple non-equilibrium lattice model which exhibits a continuous phase transition belonging to the directed percolation (DP) universality class. Each site is in either of two states, active (infected) or inactive (healthy). During the time evolution of the system, active sites can infect random neighbors with probability $p$ or recover (i.e. become inactive) with complementary probability $1-p$. After each time step (infection attempt or self-recovery) the simulation time is incremented by $\Delta t=1/N_a$, where $N_a$ is the number of active particles in the system prior to the update. As soon as the system enters the so-called absorbing state where the whole lattice is empty, the dynamic terminates.

\begin{figure}[t]
	\centering
	\includegraphics[width=\columnwidth]{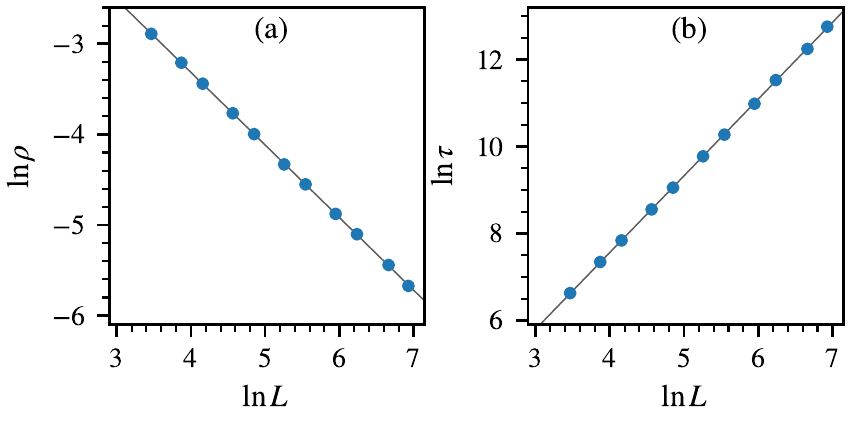}
	\caption{Quasi-stationary density (a) and lifetime (b) as a function of the linear lattice size $L$, ranging from 32 to 1024.  The slope of $\ln{\rho}$ yields $\beta/\nu$, whereas $\ln{\tau}$ yields $z$. The dots represent averages over 320 independent disorder realizations. The error bars are smaller than the symbol sizes.}
	\label{fig:qs-results}		
\end{figure}

In order to determine the critical point $p_c$ that separates the absorbing (subcritical) and active (supercritical) phase, we conduct simulations starting from a single active seed \cite{hinrichsen2000,henkel2009} on an otherwise empty \vdplus{} lattice of size $ L=6000 $, allowing for times up to $ T=10^5 $. Unwanted finite size effects are avoided by ensuring that cluster diameters remain smaller than $L$. We use 1000 independent disorder realizations with 2000 runs each. For $p_c=0.589775(3)$ we find that the average cluster size follows a power-law behavior, $\langle N_a(t)\rangle \sim t^\theta$, as expected in the case of clean universality. The uncertainty of the critical point is estimated from simulations at probabilities $p=p_c\pm \delta p$, $\delta p \ll 1$, such that the corresponding time evolution barely, but noticeably bends away from from a straight line in a double-logarithmic plot. A linear fit in the region $10^3<t<10^5$ yields an exponent $\theta=0.230(8)$, where the error stems from the uncertainty of the critical point, a result compatible with the reference value $\theta=0.2293(4)$ of the clean DP class \cite{dickman1999}.

Once the critical point is known, we can obtain the quasi-stationary (QS) density $\rho$, as well as the average lifetime of the quasi-stationary state, $\tau$, directly at $p_c$. For details on the QS method, we refer the reader to Refs.~\cite{dickman2005,oliveira2005}. As emphasized in \cite{oliveira2008b},
very long simulations times are needed to reliably detect a deviation from the universal behavior. We therefore use 320 independent realizations of the \vdplus{} lattice, ranging from $ L=32 $ to $ 1024 $, and simulate the contact process for a time of $10^8$. The measurements are taken after a generously sized relaxation period of $7\cdot 10^7$. The resulting data, shown in Fig.~\ref{fig:qs-results}, reveals that $\ln{\rho}$ and $\ln{\tau}$ follow straight lines, and linear fits yield the exponents
\begin{align}
\beta/\nu = 0.800(5), \\
z = 1.758(13), \label{z}
\end{align}
where the errors reflect both the uncertainty of the critical point and the fluctuations of the individual data points, with the latter accounting for roughly one third of the total uncertainties. Both exponents are compatible with reference values, $ \beta/\nu = 0.797(3)$ and $z=1.7674(6)$ \cite{dickman1999}. In fact, the power law dependence displayed by $\tau$ is in itself an indication of clean universal behavior, since an exponential scaling,  $\ln\tau \sim L^\psi$, is expected for uncorrelated randomness~\cite{oliveira2008b}.

\begin{figure}[t]
	\centering
	\includegraphics[width=\columnwidth]{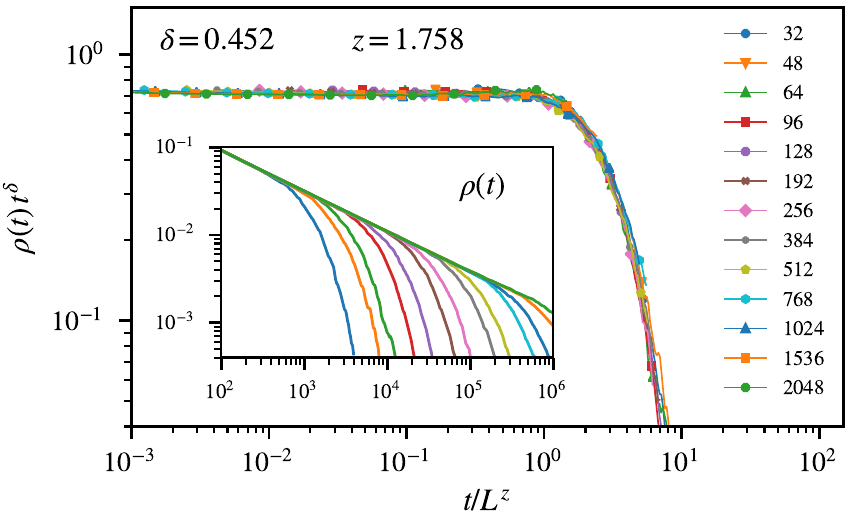}
	
	\caption{Finite size data collapse of simulations starting from a fully occupied \vdplus{} lattice at the critical point $p_c=0.589775$, using the critical exponent estimates stated in the figure. All curves are averages over 500 disorder realizations with 5 runs per realization. $L$ denotes the linear system size. The inset shows the non-rescaled density as a function of time.}
	\label{fig:collapse-fss}		
\end{figure}

In another set of simulations we measured the density of particles $\rho(t)$, starting from a fully occupied lattice. For finite-size systems at the critical point it is expected to follow 
\begin{align}
\rho(L,t) = t^{-\delta} \tilde{\rho}\left(t/L^z\right),
\end{align}
where  $ \tilde{\rho} $ is a scaling function. Performing simulations on lattices from $ L=32 $ to $ 2048 $, running for times up to $ 10^6 $ and using 500 disorder realizations with 5 runs on each, we obtain the data presented in Fig.~\ref{fig:collapse-fss}, which shows the finite-size data collapse, as well as the original measurement of $\rho(t)$ in the inset. The data collapse is performed by, using the estimate~(\ref{z}) for $z$, determining the value of $\delta$ for which the curves in Fig.~\ref{fig:collapse-fss} superpose each other, and gives us the result:
\begin{align}
\delta = 0.452(4).
\end{align}
This estimate is again compatible with the clean reference value $\delta=0.4523(10)$ from Ref.~\cite{dickman1999}.

\begin{figure}[t]
	\centering
	\includegraphics[width=\columnwidth]{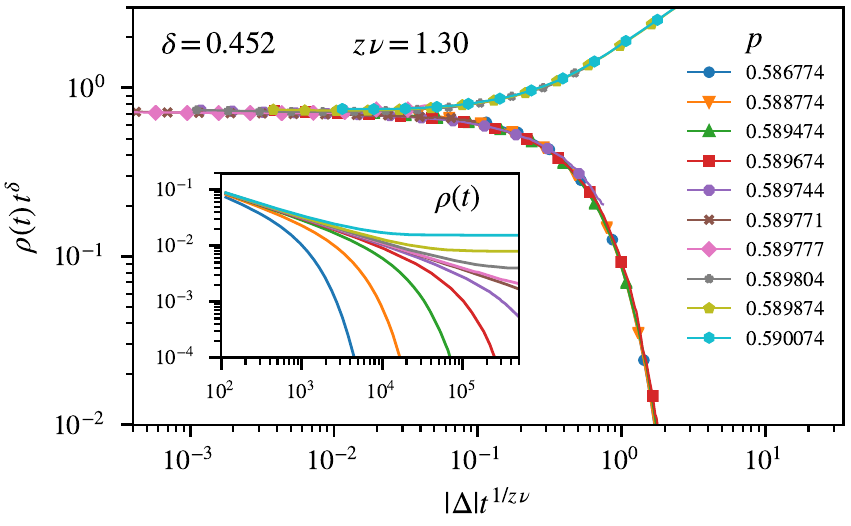}
	
	\caption{Data collapse in the off-critical region, starting from a fully occupied \vdplus{} lattice of size $L=2048$ using the critical exponent estimates given in the figure. All curves are averages over 250 disorder realizations, with 5 runs per realization. The symbol $\Delta$ denotes the distance from the critical point $p_c=0.589775$. The inset shows the non-rescaled density as a function of time. }
	\label{fig:collapse-offcritical}		
\end{figure}

We can obtain a further exponent, $\nu_\parallel = z\nu$, from off-critical simulations, using the relation
\begin{align}
	\rho(\Delta,t) = t^{-\delta} \hat{\rho}\left(\Delta t^{1/\nu_\parallel}\right),
\end{align}
where $\Delta=p-p_c$ denotes the distance from criticality and $\hat{\rho}$ is a scaling function. Using the estimate for $\delta$ obtained above, we produce the curve collapse of Fig.~\ref*{fig:collapse-offcritical} when we set
\begin{align}
	\nu_\parallel = 1.30(2),
\end{align} 
which is again compatible with the known universal value, $\nu_\parallel = 1.292(4)$ \cite{dickman1999}.

In summary, all the exponents we obtain for the two-dimensional VD$^+$ lattice in numerical simulations of the contact process, turn out to be fully compatible with the ones for the clean universality class. In particular, the data collapse plots are flawless within numerical precision.
This provides strong evidence that the phase transition on the VD$^+$ is in fact controlled by clean universal behaviour of the DP class, hence indicating that the VD$^+$ violates the HBV criterion. Clearly, regardless of our extensive numerical effort, including large lattice sizes and long simulation times, the possibility of a crossover to a non-universal behaviour for extremely long times can not be ruled out, though very large crossover times are unlikely for VD, as reasoned in~\cite{barghathi2014}.  

The results also strengthen an assumption two of us made recently~\cite{schrauth2018}: that fluctuations in the coordination number do not exclusively determine the stability of the phase transition against quenched disorder. A further study on this subject, which considers the contact process on the topologically disordered lattice with constant coordination number (CC) also introduced in~\cite{schrauth2018}, is currently in preparation. In particular, since the VD$^+$ lattice is clearly not planar (triangulations are maximal sets of non-intersecting edges~\cite{okabe2000}, thus added bonds necessarily cross existing ones), our results answer a question raised in \cite{schrauth2018}, as planarity is shown \emph{not} to be a necessary condition for stability of the phase transition against quenched spacial disorder.

\begin{acknowledgments}
We thank H.~Hinrichsen for helpful discussions. This work is part of the DFG research project Hi~744/9-1. M.S.~thanks the Studienstiftung des deutschen Volkes for financial support. F.G. thanks the DFG for funding through the SFB  1170 ``Tocotronics'' under the grant numbers Z03.
\end{acknowledgments}

\def\aj{AJ}%
\def\araa{ARA\&A}%
\def\arfm{ARFM}%
\def\apj{ApJ}%
\def\apjl{ApJ}%
\def\apjs{ApJS}%
\def\ao{Appl.~Opt.}%
\def\apss{Ap\&SS}%
\def\aap{A\&A}%
\def\aapr{A\&A~Rev.}%
\def\aaps{A\&AS}%
\def\azh{AZh}%
\def\baas{BAAS}%
\def\jrasc{JRASC}%
\def\memras{MmRAS}%
\def\mnras{MNRAS}%
\def\pra{Phys.~Rev.~A}%
\def\prb{Phys.~Rev.~B}%
\def\prc{Phys.~Rev.~C}%
\def\prd{Phys.~Rev.~D}%
\def\pre{Phys.~Rev.~E}%
\def\prl{Phys.~Rev.~Lett.}%
\def\pasp{PASP}%
\def\pasj{PASJ}%
\def\qjras{QJRAS}%
\def\skytel{S\&T}%
\def\solphys{Sol.~Phys.}%
\def\sovast{Soviet~Ast.}%
\def\ssr{Space~Sci.~Rev.}%
\def\zap{ZAp}%
\def\nat{Nature}%
\def\iaucirc{IAU~Circ.}%
\def\aplett{Astrophys.~Lett.}%
\def\apspr{Astrophys.~Space~Phys.~Res.}%
\def\bain{Bull.~Astron.~Inst.~Netherlands}%
\def\fcp{Fund.~Cosmic~Phys.}%
\def\gca{Geochim.~Cosmochim.~Acta}%
\def\grl{Geophys.~Res.~Lett.}%
\def\jcp{J.~Chem.~Phys.}%
\def\jgr{J.~Geophys.~Res.}%
\def\jqsrt{J.~Quant.~Spec.~Radiat.~Transf.}%
\def\memsai{Mem.~Soc.~Astron.~Italiana}%
\def\nphysa{Nucl.~Phys.~A}%
\def\physrep{Phys.~Rep.}%
\def\physscr{Phys.~Scr}%
\def\planss{Planet.~Space~Sci.}%
\def\procspie{Proc.~SPIE}%
\def\jpa{J.~Phys.~A: Math.~Theor.}%
\def\prep{Phys.~Rep.}%
\def\rmp{Rev.~Mod.~Phys.}%
\def\ptps{Prog.~Theor.~Phys.~Supp.}%
\def\cpc{Comput.~Phys.~Commun.}%
\def\jstate{J.~Stat.~Mech.-Theory~E.}
\def\epja{Eur.~Phys.~J.~A}%
\def\epjb{Eur.~Phys.~J.~B}%
\def\syszool{Sys.~Zool.}
\def\jpcm{J.~Phys.-Condens.~Mat.}
\def\jpssp{J.~Phys.~C:~Solid~State~Phys.}%
\def\jstat{J.~Stat.~Phys.}%
\def\zfpa{Z.~Phys.~A-Hadron.~Nucl.}
\def\plb{Phys.~Lett.~B}
\def\ijmpc{Int.~J.~Mod.~Phys.~C}
\def\fundmath{Fund.~Math.}
\def\cgp{Comput.~Graph.~Forum}
\def\physicaA{Physica A}
\def\jpcs{J.~Phys.~Conf.~Ser.}
\def\npb{Nucl.~Phys.~B}
\def\jcp{J.~Chem.~Phys.}
\def\dcg{Lect.~Notes~Comput.~Sc.}
\def\scirep{Sci.~Rep.}
\def\epjplus{Eur.~Phys.~J.~Plus}
\def\ptp{Prog.~Theor.~Phys.}%
\def\dam{Discrete~Appl.~Math.}
\def\jpcm{J.~Phys.-Condens.~Mat.}

\let\astap=\aap
\let\apjlett=\apjl
\let\apjsupp=\apjs
\let\applopt=\ao

\end{document}